# Kinetic Study of anti-HIV drugs by Thermal Decomposition Analysis: A Multilayer Artificial Neural Network Propose

B. D. L. Ferreira[1], B. C. R. Araujo[1], R. C. O. Sebastião[1]*, M. I. Yoshida[1], W. N. Mussel[1], S. L. Fialho[2], J. Barbosa[2].

[1]Departamento de Química, Universidade Federal de Minas Gerais, 31270-901, Belo Horizonte-MG, Brasil
[2]Fundação Ezequiel Dias, 30510-010, Belo Horizonte - MG, Brasil.

*Corresponding author: Sebastião, R.C.O. Tel /Fax: +553134095770
E-mail address: ritacos@ufmg.br  (R.C.O. Sebastião)

**Abstract**

Kinetic study by thermal decomposition of antiretroviral drugs, Efavirenz (EFV) and Lamivudine (3TC), usually present in the HIV cocktail, can be done by individual adjustment of the solid decomposition models. However, in some cases unacceptable errors are found using this methodology. To circumvent this problem, here is proposed to use a multilayer perceptron neural network (MLP), with an appropriate algorithm, which constitutes a linearization of the network by setting weights between the input layer and the intermediate one and the use of Kinetic models as activation functions of neurons in the hidden layer. The interconnection weights between that intermediate layer and output layer determines the contribution of each model in the overall fit of the experimental data. Thus, the decomposition is assumed to be a phenomenon that can occur following different kinetic processes. In the investigated data, the kinetic thermal decomposition process was best described by R1 and D4 model for all temperatures to EFV and 3TC, respectively. The residual error adjustment over the network is on average $10^3$ times lower for EFV and $10^2$ times lower for 3TC compared to the best individual kinetic model that describes the process. These improvements in physical adjustment allow detailed study of the process and therefore a more accurate calculation of the kinetic parameters such as the activation energy and frequency factor. It was found $E_a = 75.230$ kJ / mol and $\ln(A) = 3.2190 \times 10^{-16} s^{-1}$ for EFV and $E_a = 103.25$ kJ / mol and $\ln(A) = 2.5587 \times 10^{-3} s^{-1}$ for 3TC.

**Introduction**

Lamivudine (3TC) and Efavirenz (EFV) are antiretroviral agents, nucleoside-nucleotide inhibitors and non-nucleoside-nucleotide, respectively, of reverse transcriptase. These substances are present in HIV cocktail and are responsible for the decrease in spread of the virus through genetic recoding by enzymatic action. Thus, the HIV virus lies dormant and prevents immunosuppression of the patient. Despite the excellent results achieved in recent years, there are reports of resistance and side effects of these drugs, which led to a growing demand for less harmful analogous to patient health. Factors such as the quality and product stability must also be known, since they are important to ensure survival of the patient [1-4].

Thermogravimetry (TG) is a thermo analytic technique widely used in the pharmaceutical sciences. It allows the determination of thermal stability and the kinetic parameters through adjusting experimental isotherms using kinetic models reported in literature [4-6]. The results are used to control quality of the final products, and help in the synthetic route of new drugs [1,4,8]. However, in some instances, adjustments of experimental curves promoted by the models have unacceptable errors, providing kinetic parameters of low reliability and discredit of the used methodology. To work around this problem, we suggest a more effective methodology that uses artificial neural networks [9].

The multilayer perceptron artificial neural network (MLP) used in this study, employs an algorithm that sets the weights in the input layer according to the rate constants determined by fitting the experimental data of the kinetic models. This ensures linearization of the network using kinetic models as activation function of the neurons in the intermediate layer. The interconnection weights between that intermediate layer and the output layer determines the contribution of each model in the overall fit of the experimental data [9,10].

The MLP network is a robust mathematical method in the study of thermal decomposition, as well as considers the contribution of several models in the process, showing a significant reduction in residual error fit. This fit improvement allows detailed study of the physical process of thermal decomposition and consequently much more precise calculation of the kinetic parameters.

**Kinetic Models: Theoretical Background**

The kinetic of solid thermal decomposition is based on nucleation and growth of active nuclei present in the crystal surface. The existence of separate reactive sites related to imperfections in crystals causes an increase in the Gibbs energy of the system and hence its reactivity. Several physical models are used to describe the isothermal experimental decomposition fraction, α in function of decomposition time [5-7,11]. The physical models are given in Tab.01 and are generally classified according to the shape of the curve and its relation with the acceleration and deceleration process.

**Table 01.** Kinetic Models.

| Model | Symbol | Kinetic Equation |
|---|---|---|
| **Aceleration** | | |
| E Potencial Law | $P_n$ | $\alpha^{1/n} = kt + k_0$  $n = 2,3,4,\ldots$ |
| **Sigmoid** | | |
| Avrami-Erofeev | $A_m$ | $[-ln(1-\alpha)]^{1/m} = kt + k_0$ <br> m = 2, 3, 4... |
| Avrami-Erofeev | $A_u$ | $\ln\dfrac{\alpha}{1-\alpha} = kt + k_0$ |
| Prout-Tompkins | $A_x$ | $\ln\dfrac{\alpha}{1-\alpha} = k \ln t + k_0$  $k > 1$ |
| **Desaceleration** | | |
| **Geometric Model - contração** | | |
| Linear Contraction | $R_1$ | $1-(1-\alpha) = kt + k_0$ |
| Area Contraction | $R_2$ | $1-(1-\alpha)^{1/2} = kt + k_0$ |
| Volume Contraction | $R_3$ | $1-(1-\alpha)^{1/3} = kt + k_0$ |
| **Diffusion Model** | | |
| One Dimension | $D_1$ | $\alpha^2 = kt + k_0$ |
| Two Dimensions | $D_2$ | $(1-\alpha)\ln(1-\alpha) + \alpha = kt + k_0$ |
| Three Dimensions | $D_3$ | $\left[1-(1-\alpha)^{1/3}\right]^2 = kt + k_0$ |
| Ginstling-Brounshtein | $D_4$ | $1 - \dfrac{2\alpha}{3} - (1-\alpha)^{2/3} = kt + k_0$ |

The decomposition reaction in which the dominant stage is the acceleration step is characterized by presenting only the nucleation phenomenon. In these reactions the formation of nuclei can occur instantly or with constant nucleation rate and the exponential model is more satisfactory to describe the event. For reactions in which it is observed a chaotic nucleation followed by growth of these nuclei, the Avrami-Erofeev or Prout-Tompkins models are more suitable [6,7,11-13].

In reactions in which there is only the growth of the nuclei, the deceleration curves are most appropriate, and two phenomena are responsible for their kinetic decomposition: the contraction and diffusion. The contraction is responsible for the rapid development of the nucleus throughout the length of the crystal surfaces. Since diffusion is responsible for controlling the reaction rate, the reaction requires continuity of the transport of reactants to product layer [6,7,11-13].

The determination of the physical model in the decomposition process is crucial for the kinetic study and can be carried out by microscopy or by commercial software which employ at least three models to fit the experimental data [7]. However, these adjustments do not always show good results, since the overall decomposition process takes place in multi-stages, which requires more than one model on its description.

**Multilayer Perceptron Neural Network Methodology**

Artificial neural network consists of a mathematical algorithm based on the biological model for solving linear and nonlinear problems in several areas of science. The processing unit is an artificial neuron and its architecture is variable, being optimized according to the characteristics of the problem under study [14,15]. In the treatment of isotherms in thermal analysis, the network proved to be adequate due to the mathematical similarity between the time dependence of the decomposition fraction and the neuron impulse.

The MLP network proposed in this work is based on the algorithm originally proposed by Sebastião in 2003 [9] and 2004 [10] for the study of thermal decomposition of rhodium and rhodium (II) acetate, respectively. The network consists of three layers, one input, one output and an intermediate one. The input and output layers have only one neuron and the intermediate layer has a variable number of neurons, depending on the number of kinetic models to be considered in the process of decomposition. The first step is to determine the interconnection weights between the input layer and the intermediate layer, which is stored in a vector, $W_1$. The values of these weights are determined by the constant rates, $k$ and $k_0$ obtained from the fit of experimental data by kinetic models. The states of the neurons in the hidden layer are determined by the vector multiplication $W_1$ and time data, $t$, received by the input layer neuron.

In the second stage of the algorithm, the neurons of the intermediate layer are activated in a non-linear transformation, in which the intermediate neurons are activated by a function corresponding to a mathematical equation of the kinetic model. This step is possible because of the similarity between the shape of the kinetic curves and the usual neuron activation functions of the MLP.

In the third step the neuron on output layer receives the information of the neurons from the intermediate layer, being pondered by weights of interconnection between the intermediate layer and output. These interconnection weights are stored in another vector, $W_2$, which are obtained through a network optimization process. A network error function is determined by the difference between the values of the state of the output layer neuron $Y_{cal}$, and the experimental data of decomposed fractions of drugs, , $E = \|Y_{cal} - Y_{exp}\|_2^2$, where $Y_{cal} = W_2 f(W_1 t)$.

With this network structure is possible to calculate the contribution of kinetic models that are used in the process through a Levenberg-Marquart optimization, $W_2 = (B^T B)^{-1} B^T Y_{cal}$, where $B = f(W_1 t)$ [9,10].

Interestingly, this algorithm stands out from others because it is a robust mathematical method in the study of thermal decomposition, it uses the rate constants as fixed weights of the network enabling network convergence to a coherent physical result, in addition to considering the contribution of several models in the process. So this algorithm becomes a useful tool in the study of thermal decomposition not only of the studied systems but over to any system of interest.

**Experimental section**

The thermal decomposition experimental data for Efavirenz and Lamivudine were obtained in a thermal balance Shimadzu TGA-50H. The TGA curve, shown in Fig.01 were obtained in dynamic atmosphere of $N_2$ at rate of 100 mL/min and a heating rate of 20 ° C/min up to desired isothermal temperatures. Due to significant weight loss at 175 °C for the EFV and 225 °C for 3TC, Fig.01, these experimental curves indicate the beginning temperature of the decomposition process. Thus, the experimental isotherms, five to EFV and nine for 3TC were obtained under the same conditions of the TG, shown in Fig.02.

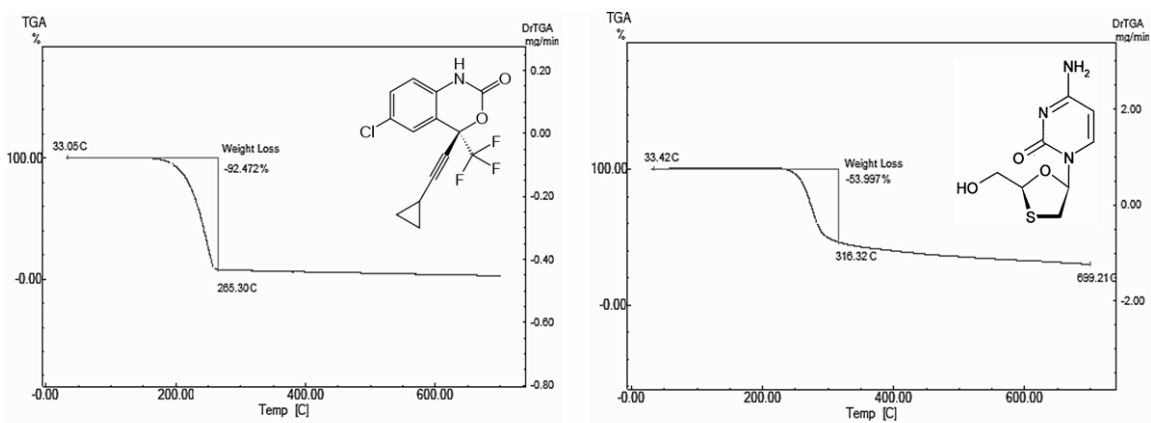
(a) (b)

**Figure 01 -** TGA and isothermal Curve for: a) Efavirenz and b) Lamivudine sample, repectively.

**Results and Discussion**

The experimental isotherms, shown in Fig.02, were used in this study to investigate the kinetic decomposition of drugs. The neural network intends to indicate the contribution of physical models that best describe the entire process. This is a very important study because it allows estimating the stability of drugs through the kinetic parameters, such as activation energy and the frequency factor.

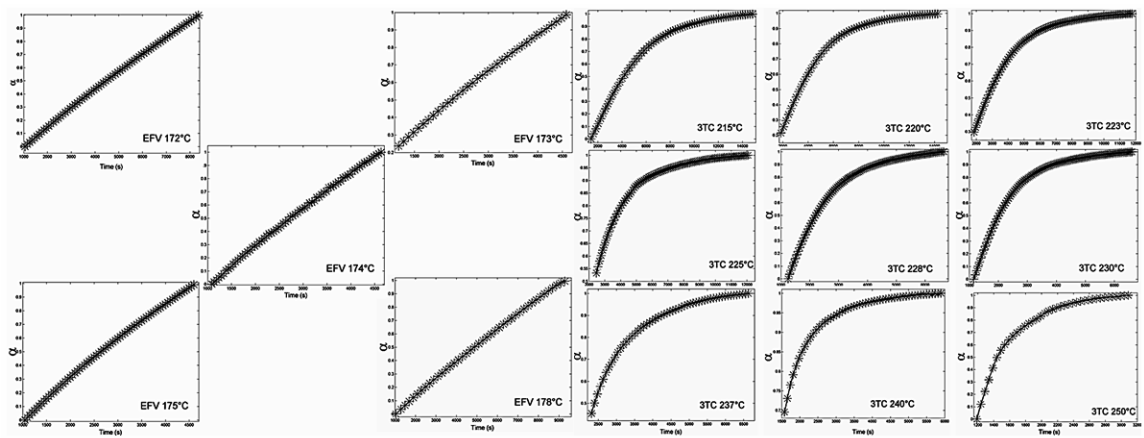
(a) (b)

**Figure 02 -** Isothermal experimental data (*) and MLP response (full line) for a) Efavirenz and b) Lamivudine sample.

In the proposed network, the neurons of the intermediate layer are activated by mathematical functions of kinetic models, described in Tab.01, as meeting the criteria for functions that are used as activation functions. Thus, the network architecture is constituted by a neuron in the input layer which receives the experimental time data; eleven neurons in the hidden layer, corresponding to the kinetic models and one neuron in the output layer, which represents the decomposition fraction of the data calculated by the network. Initially the weights of interconnection between the input layer and the intermediate layer are obtained by the constants $k$ and $k_0$, being determined by adjusting each isotherm for each physical model. The obtained constants are shown in Tab.02 for Efavirenz and Tab.03 for Lamivudine.

**Table 2.** Rate constant for kinetic models for Efavirenz at 172, 173, 174, 175 and 178 °C.

| Model | T=172°C | | T=173°C | | T=174°C | | T=175°C | | T=178°C | |
|---|---|---|---|---|---|---|---|---|---|---|
| | $k\ (10^{-4})$ | $k_0\ (10^{-1})$ | $k\ (10^{-4})$ | $k_0\ (10^{-1})$ | $k\ (10^{-4})$ | $k_0\ (10^{-1})$ | $k\ (10^{-4})$ | $k_0\ (10^{-1})$ | $k\ (10^{-4})$ | $k_0\ (10^{-1})$ |
| $D_1$ | 1.3905 | -2.9867 | 2.7305 | -3.4001 | 2.8908 | -4.6574 | 2.8474 | -4.3866 | 1.2648 | -2.9159 |
| $D_2$ | 1.1789 | -2.8530 | 2.4456 | -3.6668 | 2.4708 | -4.3164 | 2.4049 | -4.0335 | 1.0949 | -2.8834 |
| $D_3$ | 0.56359 | -1.5580 | 1.2002 | -2.1262 | 1.2142 | -2.3335 | 1.1032 | -2.0303 | 0.54763 | -1.6684 |
| $D_4$ | 0.33397 | -0.84785 | 0.70444 | -1.1266 | 0.70442 | -1.2728 | 0.67602 | -1.1728 | 0.31448 | -0.8735 |
| $R_1$ | 1.3579 | -1.1797 | 2.1952 | -0.020413 | 2.7995 | -2.7507 | 2.7578 | -2.4677 | 1.2042 | -0.96471 |
| $R_2$ | 1.1120 | -1.7184 | 2.0159 | -1.5276 | 2.3180 | -3.0659 | 2.2487 | -2.7702 | 1.0136 | -1.6588 |
| $R_3$ | 0.89941 | -1.5992 | 1.6822 | -1.6427 | 1.8886 | -2.7212 | 1.8047 | -2.4251 | 0.83089 | -1.5966 |
| $Am_4$ | 1.2172 | 3.4894 | 1.7809 | 5.0630 | 2.5383 | 2.0125 | 2.4269 | 2.4274 | 1.0849 | 3.6908 |
| $Am_2$ | 2.1923 | -1.0937 | 3.6944 | 0.078291 | 4.6183 | -3.8600 | 4.3416 | -2.9642 | 2.0054 | -0.98015 |
| AU | 9.3224 | -43.694 | 12.337 | -27.921 | 19.562 | -55.244 | 18.400 | -51.569 | 8.2830 | -41.858 |
| $F_1$ | 4.3975 | -9.8353 | 8.4718 | -11.441 | 9.5248 | -15.990 | 8.4616 | -13.142 | 4.1938 | -10.362 |

**Table 3.** Rate constant for kinetic models for Lamivudine in: 215, 220, 223, 225, 228, 230, 237, 240 and 250 °C.

| Model | T=215°C | | T=220°C | | T=223°C | | T=225°C | | T=228°C | |
|---|---|---|---|---|---|---|---|---|---|---|
| | $k\ (10^{-5})$ | $k_0\ (10^{-1})$ | $k\ (10^{-5})$ | $k_0\ (10^{-1})$ | $k\ (10^{-5})$ | $k_0\ (10^{-1})$ | $k\ (10^{-5})$ | $k_0\ (10^{-1})$ | $k\ (10^{-5})$ | $k_0\ (10^{-1})$ |
| $D_1$ | 8.1819 | -0.42727 | 16.101 | -2.7835 | 7.8487 | 2.1346 | 5.9744 | 3.9478 | 19.439 | -1.2642 |
| $D_2$ | 8.2270 | -1.4149 | 11.494 | -2.1705 | 9.1157 | 0.27727 | 7.6701 | 1.7776 | 19.779 | -2.3970 |
| $D_3$ | 5.3611 | -1.7398 | 3.7441 | -0.76461 | 3.0874 | -0.21995 | 2.7453 | 0.16877 | 13.045 | -2.4939 |
| $D_4$ | 2.5763 | -0.58625 | 2.9190 | -0.56717 | 7.4462 | -1.6855 | 7.3118 | -1.2917 | 6.2208 | -0.91339 |
| $R_1$ | 6.5424 | 2.0499 | 16.496 | -0.64888 | 5.0539 | 5.0324 | 3.5420 | 6.4577 | 15.070 | 1.6164 |
| $R_2$ | 6.8283 | 0.10401 | 11.768 | -1.0183 | 6.9969 | 1.9570 | 5.8797 | 3.1259 | 16.144 | -0.57969 |
| $R_3$ | 6.1504 | -0.49340 | 8.7932 | -0.88375 | 6.9647 | 0.70238 | 6.2260 | 1.5664 | 14.661 | -1.1802 |
| $Am_4$ | 6.9491 | 5.6184 | 11.388 | 5.2046 | 6.6833 | 7.8728 | 5.9467 | 8.7387 | 15.758 | 5.2414 |
| $Am_2$ | 1.4556 | 1.7068 | 20.505 | 1.3775 | 16.495 | 4.5917 | 15.384 | 6.1615 | 34.355 | 0.24443 |
| AU | 59.301 | -32.766 | 70.189 | -24.092 | 59.740 | -15.264 | 57.678 | -11.063 | 127.09 | -32.502 |
| $F_1$ | 38.450 | -9.5788 | 34.067 | -4.2573 | 52.929 | -8.6997 | 53.401 | -6.8405 | 93.367 | -14.808 |

| Model | T=230°C | | T=237°C | | T=240°C | | T=250°C | |
|---|---|---|---|---|---|---|---|---|
| | $k\ (10^{-5})$ | $k_0\ (10^{-1})$ | $k\ (10^{-5})$ | $k_0\ (10^{-1})$ | $k\ (10^{-5})$ | $k_0\ (10^{-1})$ | $k\ (10^{-5})$ | $k_0\ (10^{-1})$ |
| $D_1$ | 18.526 | -0.43319 | 16.249 | 0.40895 | 8.6026 | 5.7137 | 28.330 | 1.6695 |
| $D_2$ | 19.362 | -1.7117 | 19.382 | -2.0476 | 12.652 | 3.4117 | 41.872 | -2.6683 |
| $D_3$ | 1.3666 | -2.3383 | 16.207 | -3.8258 | 15.512 | -1.0218 | 47.830 | -7.7211 |
| $D_4$ | 6.2123 | -0.73792 | 6,6235 | -1.0477 | 4.9395 | 0.59692 | 16.087 | -017284 |
| $R_1$ | 14.146 | 2.3113 | 10.063 | 4.1471 | 4.7708 | 7.6389 | 15.470 | 5.4690 |
| $R_2$ | 15.792 | -0.029748 | 14.665 | 0.28661 | 10.199 | 4.2001 | 32.769 | 0.47343 |
| $R_3$ | 14.664 | -0.78426 | 14.801 | -1.0823 | 11.857 | 2.3076 | 37.243 | -2.9465 |
| $Am_4$ | 15.701 | 5.6673 | 14.086 | 6.2394 | 11.891 | 9.2581 | 36.066 | 4.3027 |
| $Am_2$ | 35.015 | 0.94189 | 35.310 | 0.25888 | 32.540 | 6.8508 | 97.383 | -6.4253 |
| AU | 131.35 | -30.636 | 128.89 | -31.421 | 130.80 | -1.583 | 380.24 | -62.301 |
| $F_1$ | 99.875 | -14.400 | 116.08 | -24.077 | 125.47 | -8.9641 | 363.29 | -57.357 |

The residual errors for the experimental data fit by the individual kinetic models are shown in Tab.04 and Tab.05 for Lamivudine and Efavirenz, respectively. Fig.02 shows the experimental data and adjustment promoted by the neural network considering all the physical models. The residual adjustment errors obtained by the network were $7.2477 \times 10^{-05}$, $2.77986 \times 10^{-05}$, $7.20793 \times 10^{-04}$, $3.59147 \times 10^{-05}$ e $7.56512 \times 10^{-05}$ for Efavirenz and $1.8196 \times 10^{-04}$, $6.0850 \times 10^{-06}$, $1.2664 \times 10^{-05}$, $4.9161 \times 10^{-05}$, $2.0487 \times 10^{-04}$, $6.1519 \times 10^{-04}$, $1.5819 \times 10^{-05}$, $1.2558 \times 10^{-04}$ and $1.5997 \times 10^{-05}$ for Lamivudine at the analyzed temperatures. As can be seen, the network residual error is on average $10^3$ times lower than the errors obtained by fitting the individual model for Efavirenz and on average, $10^2$ times lower for Lamivudine.

**Table 4.** Residual error and contribution of the individual kinetic models for Efavirenz.

| Model | T=172°C | | T=173°C | | T=174°C | |
|---|---|---|---|---|---|---|
| | Error Model (E-01) | $W_2$ | Error Model (E-01) | $W_2$ | Error Model (E-01) | $W_2$ |
| $D_1$ | 3.9003 | 0.0043513 | 2.1292 | -0.0045945 | 2.5022 | 0.084434 |
| $D_2$ | 7.8482 | -0.0089280 | 3.6514 | 0.010781 | 4.7891 | -0.078760 |
| $D_3$ | 11.118 | 0.0031939 | 6.8944 | -0.0073528 | 7.8395 | -0.039059 |
| $D_4$ | 7.7068 | 0.0070445 | 2.7202 | 0.037600 | 3.1706 | 0.064253 |
| $R_1$ | 0.094638 | 1.0000 | 0.033397 | 1.0000 | 0.080538 | 1.0000 |
| $R_2$ | 1.3782 | 0.28144 | 0.30529 | 0.18757 | 1.6228 | 0.40301 |
| $R_3$ | 7.1554 | 0.032980 | 0.82020 | 0.96547 | 19.445 | -0.0059737 |
| $Am_4$ | 24.385 | -0.074825 | 0.13000 | -0.25777 | 27.997 | -0.025679 |
| $Am_2$ | 1.2859 | -0.027863 | 0.42738 | -1.2698 | 1.4009 | -0.10393 |
| AU | 9.2968e+8 | 1.1023e-5 | 0.38297 | 0.74349 | 3.0653e+8 | 8.2146e-6 |
| $F_1$ | 28.597 | -0.022578 | 5.0257 | -0.23876 | 30.002 | -0.026513 |

| Model | T=175°C | | T=178°C | |
|---|---|---|---|---|
| | Error Model (E-01) | $W_2$ | Error Model (E-01) | $W_2$ |
| $D_1$ | 2.4952 | -0.016977 | 1.5197 | -0.010291 |
| $D_2$ | 5.2300 | -0.0046009 | 3.1312 | 0.0091536 |
| $D_3$ | 7.1922 | -0.0061368 | 5.1211 | 0.022280 |
| $D_4$ | 4.5057 | -0.037169 | 2.5261 | -0.036745 |
| $R_1$ | 0.20605 | 1.0000 | 0.050071 | 1.0000 |
| $R_2$ | 0.83371 | 0.88197 | 0.99839 | 0.17944 |
| $R_3$ | 3.5171 | -0.21743 | 2.6809 | -0.85077 |
| $Am_4$ | 36.982 | -0.23166 | 24.695 | -0.058278 |
| $Am_2$ | 0.79779 | 0.070089 | 0.84975 | 0.056814 |
| AU | 8.7290 e+22 | 4.5323e-5 | 8.3116 e+08 | 9.4700 e-6 |
| $F_1$ | 17.691 | 0.014141 | 23.427 | 0.085499 |

**Table 5**. Residual error and contribution of the individual kinetic models for Lamivudine.

| Model | T=215°C Error Model (E-01) | W₂ | T=220°C Error Model (E-01) | W₂ | T=223°C Error Model (E-01) | W₂ |
|---|---|---|---|---|---|---|
| $D_1$ | 3.1541 | 0.099235 | 8.6623 | -0.35874 | 4.7509 | 0.10398 |
| $D_2$ | 0.55410 | 0.023117 | 1.8425 | 0.11141 | 1.2918 | 0.01164 |
| $D_3$ | 0.31042 | -0.050424 | 2.1865 | -0.0035124 | 0.049773 | 0.40284 |
| $D_4$ | 0.017511 | 1.000 | 0.014611 | 1.0000 | 0.0033675 | 1.0000 |
| $R_1$ | 14.482 | -0.076304 | 4.5354 | 0.37675 | 2.9443 | -0.080521 |
| $R_2$ | 2.0352 | -0.058335 | 1.7463 | 0.014565 | 0.93387 | -0.36162 |
| $R_3$ | 0.31826 | -0.091536 | 0.28740 | 0.021553 | 0.27900 | -0.09965 |
| $Am_4$ | 23.216 | 0.0093202 | 0.90609 | 0.0010881 | 0.35953 | 0.17441 |
| $Am_2$ | 1.4860 | 0.25187 | 0.47310 | -0.0025675 | 0.67921 | 0.14802 |
| AU | 9.3971e+07 | -7.5230e-06 | 0.69098 | -0.020124 | 0.26558 | -0.025782 |
| $F_1$ | 14.949 | 0.010348 | 0.13451 | 0.051401 | 1.4353 | -0.036678 |

| Model | T=225°C Error Model (E-01) | W₂ | T=228°C Error Model (E-01) | W₂ | T=230°C Error Model (E-01) | W₂ |
|---|---|---|---|---|---|---|
| $D_1$ | 5.6300 | 0.097578 | 6.0776 | -0.076464 | 9.8873 | 0.068540 |
| $D_2$ | 1.5627 | -0.10840 | 2.3398 | 0.11632 | 2.4460 | -0.19158 |
| $D_3$ | 0.043316 | 0.24657 | 0.22878 | -0.19750 | 0.089087 | 0.37525 |
| $D_4$ | 0.0083716 | 1.0000 | 0.027051 | 1.0000 | 0.047886 | 1.0000 |
| $R_1$ | 4.2990 | -0.051241 | 27.280 | -0.0073638 | 14.517 | -0.048133 |
| $R_2$ | 1.1797 | -0.13854 | 7.8280 | 0.00092622 | 2.9419 | -0.055792 |
| $R_3$ | 0.34217 | 0.11220 | 0.73475 | -0.13476 | 0.68847 | 0.15205 |
| $Am_4$ | 0.74771 | -0.052142 | 9.6586 | -0.11614 | 6.9674 | -0.013417 |
| $Am_2$ | 1.5049 | 0.10484 | 7.0207 | 0.17907 | 10.265 | 0.039763 |
| AU | 0.90581 | 0.0031334 | 1.0100 e+21 | 1.8086e-5 | 1.0741 e+21 | 6.8277 e-06 |
| $F_1$ | 2.5999 | -0.021493 | 6.0776 | -0.017810 | 7.7944 | -0.025135 |

| Model | T=237°C Error Model (E-01) | W₂ | T=240°C Error Model (E-01) | W₂ | T=250°C Error Model (E-01) | W₂ |
|---|---|---|---|---|---|---|
| $D_1$ | 1.5225 | 0.271776 | 8.7199 | -1.0059 | 0.32909 | -0.68175 |
| $D_2$ | 0.45499 | 0.023793 | 3.2909 | 0.54610 | 0.11841 | 0.65237 |
| $D_3$ | 0.054316 | -0.048536 | 1.0901 | -0.095015 | 0.083541 | -0.063357 |
| $D_4$ | 0.032728 | 1.0000 | 0.78882 | 1.0000 | 0.0087238 | 1.0000 |
| $R_1$ | 5.6229 | -0.15520 | 10.436 | 0.47464 | 0.28947 | 0.42798 |
| $R_2$ | 0.85334 | -0.42502 | 6.3306 | -0.017704 | 0.067763 | -0.45121 |
| $R_3$ | 0.33777 | 0.39165 | 3.1921 | 0.089745 | 0.015999 | 0.39288 |
| $Am_4$ | 2.1074 | -0.044044 | 5.4915 | -0.13848 | 0.034821 | -0.44232 |
| $Am_2$ | 0.80278 | 0.0043719 | 5.7944 | 0.030036 | 0.090555 | -0.11428 |
| AU | 6.1319 | 0.078353 | 8.9017 | 0.0062334 | 0.27567 | 0.20390 |
| $F_1$ | 0.78544 | 0.042629 | 1.7243 | -0.036222 | 0.63328 | 0.03357 |

This decrease in the residual error if compared with individual adjustment promoted by the models can be explained based on the correction in the mathematical function of the kinetic models. The degree of freedom added to the models through the vector elements $W_2$ is an important correction factor. For instance, in the decomposition sigmoid model, for a long time, the decomposition fraction must not be unitary and residue must be taken into account, being represented by this vector element. The values of contribution for each model, $W_2$ vector elements, in the description of the Efavirenz and Lamivudine thermal decomposition at each temperature are shown in Tab.04 and 05, respectively. These values were normalized due to the higher contribution at each temperature.

Through the analysis of the results it is possible to determine the model or set of models that best describe the thermal decomposition phenomenon of drugs present in the anti-HIV cocktail under study. Fig.03 shows a representation of the models that had lower residual error and consequently higher contributions to the fit of the experimental curves. As can be seen in Fig.03, the model of one-dimensional contraction, R1 and diffusion model Ginstling-Brounshtein, D4, are the models that best describe the decomposition process for EFV and 3TC respectively.

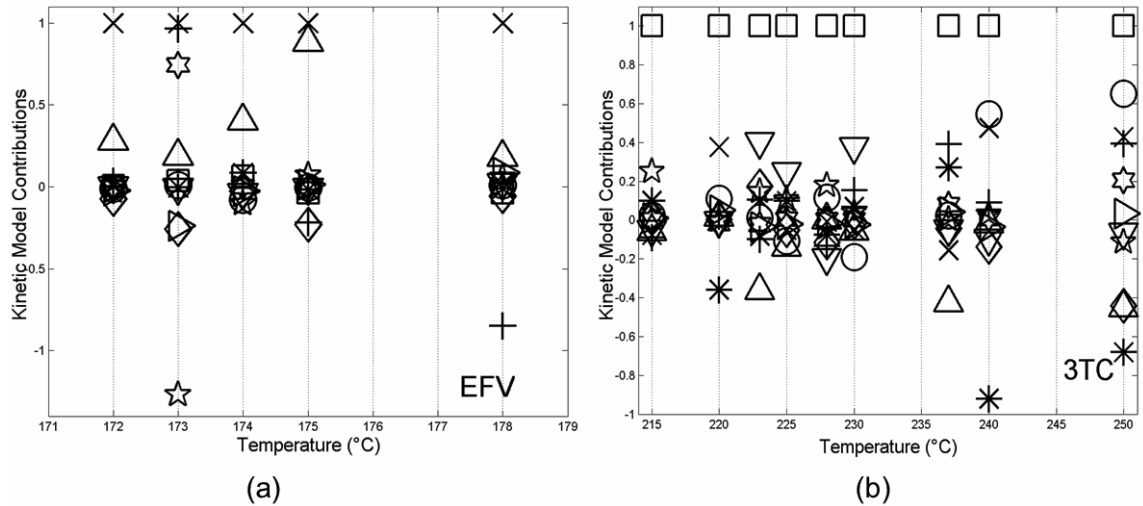

**Figure 03** – Kinetic model contributions: $D_1$ (∗), $D_2$ (O), $D_3$ (∇), $D_4$ (□), $R_1$ (×), $R_2$ (Δ), $R_3$ (+), $Am_4$ (◊), $Am_2$ (○), $Au$ (☆) and $F_1$ (▷) for a) Efavirenz and b) Lamivudine sample.

Based on these results, two further networks were proposed considering only R1 models for Efavirenz and D4 to Lamivudine. Thus, only one neuron in the hidden layer is used, but a degree of freedom is added to the kinetic model function with the weight adjustment of the interconnection weight between the intermediate layer and the output layer. The Fig.04 shows the residual error of the two networks. As can be seen, the network error is much smaller at all studied temperatures.

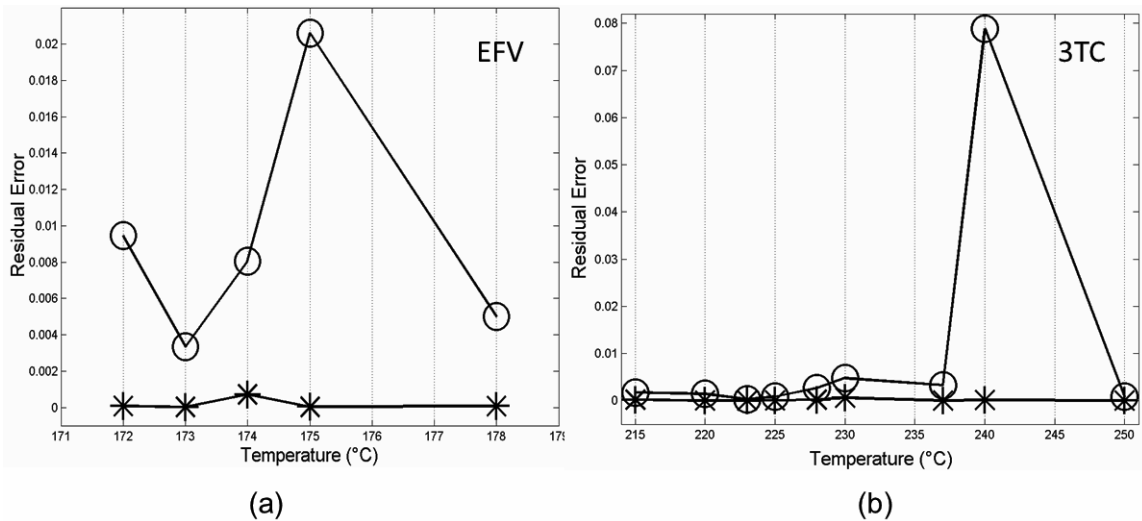

**Figure 04** – MLP Neural network residual errors (∗) with a) only R1 model (O) for Efvarirenz and b) only D4 model (O) for Lamivudine.

The activation energy, $E_a$, and the frequency factor, $\ln(A)$ were determined using the rate constants obtained by the models that had lower error in experimental data adjustment. Considering the Arhenius theory, the matched values are $E_a = 75.230$ kJ/mol and $\ln(A) = 3.2190e^{-16}s^{-1}$ for EFV and $E_a = 103.25$ kJ/mol and $\ln(A) = 2.5587e^{+3}s^{-1}$ to 3TC.

**Conclusion**

The thermal decomposition process of antiretroviral drugs Efavirenz and Lamivudine, present in the anti-HIV cocktail, is studied in this work using artificial neural network. Since the solid decomposition models show similar curves to the nervous impulse, two MLP networks, one for each substance have been proposed using the eleven physical models as activation functions in the hidden layer. This methodology is possible by the setting the interconnection weights between the input layer and the intermediate layer with the values of the calculated rate constants for each model.

The proposed algorithm makes the neural network not lose chemical information during its optimization process and the network becomes a powerful mathematical tool for treating thermal decomposition phenomenon of substances, since their residual errors are much smaller when compared to adjustments of the individual models separately. Furthermore, with this approach it is possible to calculate the individual contribution of each model, and assuming the process as a combination of models in describing the experimental data. Another correction proposed by the neural network is to consider the correct asymptotic value of the decomposition fraction. Thus, there is a significant reduction in residual error adjustment by the network. The results of this study confirm the superiority of the neural network for the study of decomposition for both drugs, which allow detailed study of the physical process and therefore the more accurate calculation of the kinetic parameters.

The described methodology is not restricted to the Lamivudine and Efavirenz drugs, suggesting a powerful routine method for kinetic studies of the thermal decomposition process. For applications in the pharmaceutical area, this model can be used to predict the thermal decomposition and stability of the drugs to assess the shelf-life time.

**Acknowledgements**

The authors would like to thank CNPq and PRPq/UFMG for financial support.